\documentclass[doublespace,12pt]{article}
\usepackage{times}
\usepackage{mathptm}
\usepackage{graphicx}
\usepackage{anysize}
\marginsize{2cm}{2cm}{2cm}{3cm}
\title{Gate-Level Simulation of Quantum Circuits\thanks{
        This work was partially supported by the DARPA QuIST program.
        The views and conclusions contained herein are those of the authors
        and should not be interpreted as necessarily representing official
        policies of endorsements, either expressed or implied, of the
        Defense Advanced Research Projects Agency (DARPA), the Air Force
        Research Laboratory, or the U.S. Government.
  }
 }

 \author{George F. Viamontes, Manoj Rajagopalan, 
         Igor L. Markov and John P. Hayes \\
         \small University of Michigan, 
                 Advanced Computer Architecture Laboratory, 
                 Ann Arbor, MI 48109-2122 \\
         {\tt \small \{gviamont,rmanoj,imarkov,jhayes\}@eecs.umich.edu}
       }

\begin{document}

\maketitle
\pagestyle{empty}
\thispagestyle{empty}

\abstract{
  
  While thousands of experimental physicists and chemists are
  currently trying to build scalable quantum computers, it appears
  that simulation of quantum computation will be at least as 
  critical as circuit simulation in classical VLSI design.
  However, since the work of Richard Feynman in the early 1980s
  little progress was made in practical quantum simulation.
  Most researchers focused on polynomial-time simulation
  of restricted types of quantum circuits that fall short
  of the full power of quantum computation \cite{Gottesman98}.

  Simulating quantum computing devices and useful quantum algorithms
  on classical hardware now requires excessive computational resources,
  making many important simulation tasks infeasible. In this work we 
  propose a new technique for gate-level simulation of quantum circuits
  which greatly reduces the difficulty and cost of such simulations. 
  The proposed technique is implemented in a simulation tool called the
  Quantum Information Decision Diagram (QuIDD) and evaluated 
  by simulating Grover's quantum search algorithm \cite{grover}.
  The back-end of our package, QuIDD Pro, is based on Binary Decision Diagrams,
  well-known for their ability to efficiently represent many seemingly 
  intractable combinatorial structures. 
  This reliance on a well-established area of research allows us to take
  advantage of existing software for BDD manipulation and achieve unparalleled
  empirical results for quantum simulation. 

}

\section{Introduction}

In the last decade, a revolutionary computing paradigm has emerged
that, unlike conventional ones such as the von Neumann model, is based
on quantum mechanics rather than classical physics
\cite{NielsenC2000}. Quantum computers can, in principle, solve some
hitherto intractable problems including factorization of large
numbers, a central issue in secure data encryption. While the state of
the art in building quantum devices is still in its infancy,
significant progress has been made recently. For example, IBM has
announced \cite{NYTimes2001} an operational quantum circuit that
successfully factored 15 into 3 and 5 using Nuclear Magnetic Resonance
(NMR) technology. Quantum circuits have also been recognized as
necessary infrastructure to support secure quantum communication,
quantum cryptography, and precise measurement.

In addition to the IBM device, somewhat smaller operational circuits
have been implemented using entirely unrelated technologies, including
ion traps, electrons floating on liquid helium, quantized currents in
super-conductors and polarized photons. While it is not clear which
technologies will ultimately result in practical quantum circuits, a
number of fundamentally valuable design and test questions can be
addressed in technology-independent ways using automated techniques.
Our work proposes such a technique and shows its practical benefits.

Automated simulation is one of most fundamental
aspects in the design and test of classical computing systems.
We only mention two reasons for this.
First, finding design faults prior to manufacturing is a major cost-saving 
measure. Simulation allows one to evaluate and compare competing designs
when such comparisons cannot be made analytically.
Second, simulation in many cases leads to a better understanding
of given designs (e.g., allows one to find the most appropriate clock 
frequency and identify critical paths sensitizable by realistic inputs).
Hundreds of simulation tools have been developed in the last 40 years
both in the academia and commercially. They span a broad range of
applications from circuit-physics simulations ({\tt Spice}) at the
level of individual transistors and wires to architectural simulations
of microprocessors ({\tt SimpleScalar}). Particularly large systems,
e.g., recent microprocessors, can be simulated in great detail
on specialized hardware, typically large sets of FPGA chips. 
Recent trends in Electronic Design
Automation are further expanding the range of automated simulation to
(i) symbolic simulation of high-level programs in Verilog or even C++,
and (ii) field-equation solvers that model high-performance
clock distribution networks.

  Simulation of quantum computation appears at least as important
 as classical simulation, but faces additional objectives and 
 additional obstacles. Given that Boolean logic and common intuition
 are insufficient to reason about quantum computation, automated 
 simulation may be important to designers of even small (10-20 qubits)
 quantum computers. Additionally, simulation may be handy in algorithm
 design. In classical experimental algorithmics, the performance of
 new heuristics is tested by implementing them on over-the-counter
 PCs and workstations. Indeed, a number of practically useful heuristics,
 such as Kernighan-Lin for graph partitioning and Lin-Kernighan for the Traveling Salesman Problem still
 defy comprehensive theoretical analysis. Since quantum computing hardware
 is not as commonly available, simulation tools are needed to support
 research on quantum algorithms whose empirical performance cannot be 
 described by provable results. We also point out that simulation-driven
 research is already common in computer architecture, where physically
 manufacturing a new microprocessor to evaluate a new architecture is
 ruled out by cost considerations. 

   As early as in 1980s Richard Feynman observed that simulating quantum
 processes on classical hardware seems to require super-polynomial
 (in the number of qubits) memory and time.
 Subsequent work \cite{Gottesman98} identified a number of special-case 
 quantum circuits for which tailor-made simulation techniques require
 only polynomial-sized memory and polynomial runtime.
 However, as noted in \cite{Gottesman98}, these ``restricted types 
 of quantum circuits fall short of the full power of quantum computation''.
 Thus, in cases of major interest --- Shor's and Grover's algorithms ---
 quantum simulation is still performed with straightforward
 linear-algebraic tools and requires astronomic resources.
 To this end, a recent work \cite{OuchiFH2002} used a 1.3 million-gate
 FPGA device from Altera running at 30MHz to simulate an 8-qubit 
 quantum circuit. Increasing the size of the simulation to 9 qubits
 can double the number of gates used.
 
 The approach pursued in our work is to improve asymptotic time and memory
 complexity of quantum simulations in cases where quantum operators involved
 exhibit significant structure. We observe that array-based representations
 of matrices and vectors used by major linear-algebra packages tend to have
 the same space complexity regardless of the values stored. Indeed, sparse
 matrix and vector storage is of limited use in quantum computing because
 tensor products of Hadamard matrices do not have any zero elements.
 Therefore we use graph-based data structures for matrices and vectors
 that essentially perform data compression and facilitate linear algebra
 operations in compressed form. While this approach does not improve
 abstract worst-case complexity, it achieves significant speed-ups
 and memory savings in important special cases.

The remaining part of the paper is organized as follows.
Section \ref{sec:backg} provides the necessary background
on quantum computing. In Section \ref{sec:theory}, we describe
the theoretical framework for our technique.  Empirical results
and complexity analyses are given in Section \ref{sec:exp}.
Finally, Section \ref{sec:conclusions} concludes 
with some final thoughts and avenues for future research.

\section{Background}
\label{sec:backg}

Below we cover the necessary background in quantum computing
\cite{NielsenC2000} and simulation tools for quantum computation
\cite{Wallace2000}.


\subsection{Quantum Computation}  
In modern computers (referred to as 'classical' to distinguish them
from their quantum counterparts) binary information is stored in a bit
that is physically a voltage signal in a solid-state electronic
circuit. Mathematically, a bit is represented as a boolean value or
variable. In the quantum domain, binary information is stored in a
quantum state such as the polarization (horizontal/vertical) of a
photon or the spin (up/down) of an electron or atomic nucleus. Unlike
a classical bit, a quantum bit or {\it qubit} can exist in a
superposition of its classical binary states that is disturbed, or in
most cases, destroyed by any external stimulus (typically in a
measurement operation). Using Dirac notation, the quantum states
corresponding to the classical logic zero and one are denoted as
$|0\rangle$ and $|1\rangle$, respectively.  \cite{NielsenC2000}.
However, unlike a classical bit, a qubit can store zero and one
simultaneously, using values represented by 2-element vectors of the
form, $\alpha |0 \rangle + \beta |1 \rangle$ where $\alpha$ and
$\beta$ are complex numbers and $| \alpha |^2 + | \beta |^2 = 1$. A
fundamental postulate of quantum mechanics dictates that these
individual state vectors can be combined, via the tensor product, with
other state vectors \cite{NielsenC2000}. These tensored qubit states
provide a kind of massive parallelism since superposition allows an
$n$-qubit state $ |\Psi \rangle = \sum^{2^n-1}_{i=0} c_i
|b_{i,n-1}b_{i,n-2},\ldots,b_{i,0} \rangle$ to store $2^n$ binary
numbers simultaneously.  Each $c_i$ above is a complex number (like
$\alpha$ and $\beta$), such that $\sum^{2^n-1}_{i=0} |c_i|^2 = 1$, and
$b_{i,n-1}b_{i,n-2},\ldots,b_{i,0}$ is the binary expansion of the
number $i$.  For example, when $n=2$, $|\Psi \rangle = c_0 |00 \rangle
+ c_1 |01 \rangle + c_2 |10 \rangle + c_3 |11 \rangle$.
 
The behavior of quantum circuits is governed by quantum mechanics, and
fundamentally differs from that of classical circuits. Variables
(signal states) are qubit vectors. Gates are linear operators over a
Hilbert space, and can be represented by unitary matrices. When a gate
operation is applied to a quantum state, the resulting state can be
computed by evaluating the corresponding matrix-vector product. Thus,
logic circuits constructed with these components perform
linear-algebraic, and as a result, are subject to the following
unavoidable quantum mechanical {\em design constraints}.

\begin{itemize}
\vspace{-1.5mm}
 \item{Gates must be reversible (information-lossless) and have the same
       number of inputs and outputs}
\vspace{-1mm}
 \item{The cloning of states in superposition is impossible}
\vspace{-1mm}
\item{The measurement of quantum states is nondeterministic: the
    resulting state is the result of probabilistic collapse and this
    transformation is irreversible.}
\vspace{-1mm}
\item{The main limiting factor in classical simulation of quantum
    circuits is the number of qubits since its memory space grows
    exponentially with it.  This parameter corresponds to the number
    of signal ``wires'' in circuit diagrams.  }  \vspace{-1mm}
\vspace{-1mm}
\item{Quantum states decay or ``decohere'' due to interaction with the
    environment (which could even be vacuum). Decoherence time limits
    the number of faultless gate operations that a quantum circuit can perform.}
\vspace{-1.5mm}
\end{itemize}

 \begin{figure}
 \begin{center} 
 $
              \left[ 
                \begin{array}{cc} 
                    \frac{1}{\sqrt{2}} & \frac{1}{\sqrt{2}} \\
                    \frac{1}{\sqrt{2}} & -\frac{1}{\sqrt{2}} \\
                \end{array}
              \right]
              \otimes
              \left[ 
                \begin{array}{cc} 
                    \frac{1}{\sqrt{2}} & \frac{1}{\sqrt{2}} \\
                    \frac{1}{\sqrt{2}} & -\frac{1}{\sqrt{2}} \\
                \end{array}
              \right]
              =
              \left[ 
                \begin{array}{cccc} 
                  \frac{1}{2} & \frac{1}{2} & \frac{1}{2} & \frac{1}{2} \\
                  \frac{1}{2} &-\frac{1}{2} & \frac{1}{2} &-\frac{1}{2} \\
                  \frac{1}{2} & \frac{1}{2} &-\frac{1}{2} &-\frac{1}{2} \\
                  \frac{1}{2} &-\frac{1}{2} &-\frac{1}{2} & \frac{1}{2} \\
                \end{array}
              \right]
 $
 \parbox{9cm}
 {
  \caption{ \label{fig:Hadamard} 
    1-qubit 2-by-2 Hadamard matrix and its 4-by-4 tensor-square (a
    $2$-qubit matrix).  } 
 }
 \end{center}
 \vspace{-5mm}
 \end{figure}
 
 Because quantum mechanics is counter-intuitive, quantum states
 fragile, and technology expensive, computer-aided design and
 verification become indispensable for even small quantum computing
 systems as the state-of-the-art in hardware technology supports only
 circuits of up to 9 qubits. Simulation is critical for early
 evaluation of implementation approaches and prototypes yet remains a
 significant challenge since the pioneering work by Richard Feynman in
 the early 1980's.  Feynman argued \cite{FeynmanOnC, FeynmanLC} that
 such simulation would be impractical because of its exponential
 demand on computational resources. Our work makes it possible to
 efficiently simulate quantum circuits on classical computers. The
 potential contribution of quantum computers to simulating quantum
 mechanical phenomena has been recognized however \cite{Zalka96}.


\subsection{Quantum Simulation on Classical Computers}
A number of ``programming environments'' for quantum computing were 
proposed recently\footnote{Examples include the QCL language
({\tt \small http://tph.tuwien.ac.at/\~{}oemer/qcl.html}),
 Quantum Fog ({\tt \small  http://www.ar-tiste.com/}) and 
Open Qubit (see {\tt \small http://www.ennui.net/\~{}quantum/}). }
that are mostly front-ends to quantum circuit simulators.
This separation is similar to what is common in classical simulation.
However, quantum back-end simulators are currently based on numerical linear
algebra software, require super-polynomial memory and are limited by the number
of qubits the available RAM can accommodate.  Such back-end simulators would
benefit immensely from techniques that support efficient linear-algebraic
operations on compressed arguments.

As described earlier, quantum gates are represented as matrices and
applied to quantum states using matrix-vector multiplication. A direct
simulation approach only requires straightforward linear algebra
operations and can be quickly implemented with interactive tools such
as MATLAB (commercial; see {\tt \small http://www.mathworks.com/}) and
Octave (open-source, available at {\tt \small http://www.octave.org/}) or with
standard, high-performance compiled libraries such as BLAS/LAPACK in
Fortran 77
(available at {\tt http://www.netlib.org/lapack/}) or Blitz++ in C++ 
\cite{blitz}.
The main problem with these general approaches is that quantum states
and gate matrices must be stored explicitly, and thus require
exponential memory. For example, $n=20$ qubits entails a $2^{20}\times
2^{20}$ complex-valued matrix, whose storage is well beyond the memory
available in modern computers. Clever improvements to straightforward
linear algebra include performing $2\times2$-gate operations on state-vectors
without explicitly computing $2^{20}\times 2^{20}$-matrix gate representations.
We use such improvements in our experiments with Blitz++, 
yet Blitz++ underperforms graph-based techniques described in the next
session.

More sophisticated simulation techniques, e.g., MATLAB's ``packed''
representation, include data compression.  It is difficult, though, to
perform matrix-vector multiplication without decompressing the
operands. Another interesting approach was explored by Greve
\cite{shornuf} in his program ({\it SHORNUF}) to simulate Shor's
algorithm \cite{Shor97} using Binary Decision Diagrams(BDDs).  {\it
  SHORNUF} uses one decision node to represent the probability
amplitudes of each qubit, allowing only two states.  Because of this,
amplitudes are restricted to $1/\sqrt{2}$ and the compression factor
is rather limited. Although Greve's BDD representation isn't scalable
and doesn't result in compression improvements for arbitrary quantum
circuits, the idea of applying a BDD-\textit{like} structure has
considerable merit in the quantum domain. In the following sections we
review the relevant properties of BDDs and explain operations on
quantum states in compressed form.

\section{QuIDD Theory}
\label{sec:theory}

This section explains how vectors and matrices in quantum computing lend
themselves to the QuIDD representation and explores how to perform
linear-algebraic operations with QuIDDs.

The idea of a Quantum Information Decision Diagram (QuIDD) was born
out of the observation that vectors and matrices which arise in quantum
computing exhibit a lot of structure. More complex operators
are obtained from the
tensor product of simpler matrices and continue to exhibit common
substructures (as will be discussed shortly). 
Graphs are a natural choice for capturing such
patterns.

\subsection{BDD Compression} Reduced Ordered Binary Decision Diagrams
(ROBDDs), and operations for manipulating them were originally
developed by Bryant \cite{bryant} to handle large Boolean functions
efficiently. A BDD is a {\em directed acyclic graph} (DAG) with up to
two outgoing edges per node, labeled "then" and "else".  A DAG can
encode exponentially many directed paths, leading to a powerful data
structure for representing and manipulating Boolean functions.
Algorithms that perform operations on BDDs are typically recursive
traversals of directed acyclic graphs.  While not improving worst-case
asymptotics, in practice BDDs achieve exponential space compression
and run-time improvements by exploiting the structure of functions
that arise in digital logic design.


Our work begins with the observation that many quantum simulations
contain frequently-repeated patterns in the form of arithmetic
expressions that are evaluated over and over.  We attempt to improve
runtime and memory consumption by automatically capturing and reusing
common sub-expressions in the average case.  We describe algorithms
and data structures that empirically achieve exponential improvements
in both processing time and memory requirements when simulating
standard quantum algorithms on classical computers.

We observe that Multi-Terminal Binary Decision Diagrams (MTBDDs)
\cite{Clarke96} and Algebraic Decision Diagrams (ADDs) \cite{Bahar93}
conveniently provide the type of compression that we need --- with
integrated linear-algebraic operations that do not require
decompression. In this paper, we define a compressed representation of
complex-valued matrices and vectors, called the Quantum Information
Decision Diagram (QuIDD). The underlying structure of a QuIDD can be
either an MTBDD \cite{Clarke96} or an ADD \cite{Bahar93}. The primary
difference between an MTBDD and an ADD is that MTBDDs were originally
defined to use integer terminals whereas ADDs offer any type of
numeric terminal.  Though in practice, packages that support ADD, such as
CUDD \cite{cudd}, do not contain interface for complex-valued terminals.
As a result, our implementation of QuIDDs, a package
called QuIDD Pro, uses the terminals as array indices that map to a
list of complex numbers. Although MTBDD terminals are equally capable
of serving as array indices, we are using the CUDD package \cite{cudd}
 rather than the MTBDD package CMUBDD \cite{cmubdd}. This is because CUDD
offers a diverse set of operations that manipulate ADDs as matrices. For the
remainder of this paper, we simply refer to ADDs, but note that MTBDDs can be
substituted without loss of generality.

QuIDDs achieve significant compression when applied to qubit states
and operators encountered in gate-level descriptions of quantum
circuits.  Space and run-time complexities of our simulations of
$n$-qubit systems range from $O(1)$ to $O(2^{n})$, but the worst case
is not typical.  Moreover, even in the worst case, our simulations are
asymptotically as fast as using straightforward computational linear
algebra, e.g., MATLAB. Our empirical measurements demonstrate that
QuIDDs lead to significantly faster simulations of quantum
computation. In our experiments, quantum computation is simulated via
matrix-vector multiplication and tensor product operations performed
directly on QuIDDs. In the next section we show how vectors and
matrices map to the QuIDD structure and explore how to construct
linear-algebraic operations with QuIDDs.

\begin{figure}[t]
\begin{center}
  \includegraphics[width=11cm]{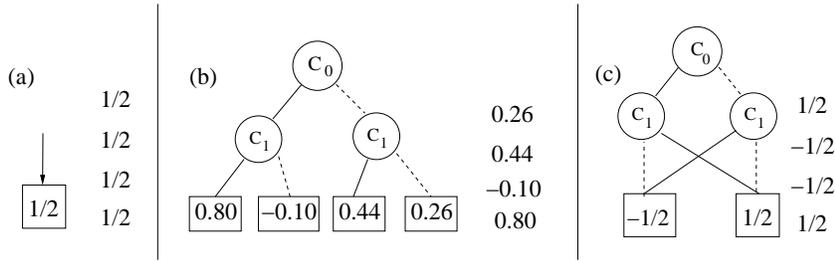}

  \parbox{15cm}{\caption{\label{fig:compress} QuIDD examples
      illustrating (a) best, (b) worst, and (c) mid-range
      complexity.}}

\end{center}
\vspace{-5mm}
\end{figure}



\subsection{Vectors and Matrices}
Figure \ref{fig:compress} shows the QuIDD structure for a few 2-qubit
state vectors. With the binary indexing of the vector elements shown
above, we define the variable nodes of a QuIDD to correspond to
decisions on index variables just as they are defined for MTBDDs and
ADDs \cite{Clarke96, Bahar93}. For example, traversing the {\it then}
edge (solid line) of node $C_0$ in Figure \ref{fig:compress}(c) is
equivalent to assigning the value $1$ to the first binary digit of the
vector index.  Similarly, traversing the {\it else} edge (dotted line)
of node $C_1$ in the same figure is equivalent to assigning the value
$0$ to the second binary digit of the vector index. It is easy to see
that given these choices on the variable index, we arrive at the
terminal node $-1/2$, which is precisely the value at index $10$ in
the explicit vector representation.


QuIDD representations of matrices extend those of vectors by adding a
second type of variable node. To motivate this, consider the following
matrix with binary row and column indices:

\begin{center} 
  $
  \begin{array} {cc}
    \begin{array} {c}
      00 \\ 01 \\ 10 \\ 11
    \end{array} &
    \left[
      \begin{array} {cccc}
        1/2 & 1/2 & 1/2 & 1/2 \\
        1/2 & -1/2 & 1/2 & -1/2 \\
        1/2 & 1/2 & -1/2 & -1/2 \\
        1/2 & -1/2 & -1/2 & 1/2
      \end{array}
    \right] \\ &
    \begin{array} {cccc}
      00~~~~~ & 01~~~~~ & 10~~~~~ & 11
    \end{array}
  \end{array}
$
\end{center}

\begin{figure}[th]
\begin{center}
  \includegraphics[width=8cm]{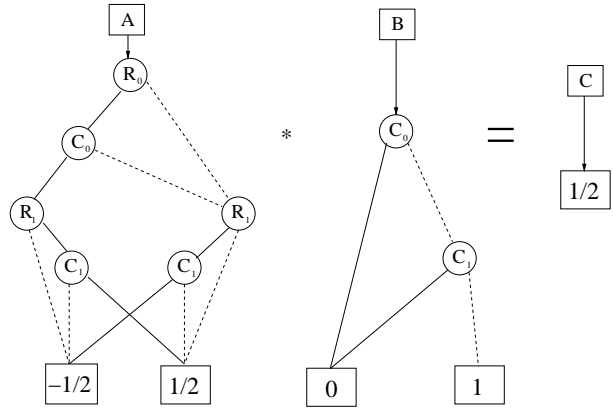}

\parbox{14cm}{\caption{\label{fig:mmult}
  Sample matrix-vector multiplication with QuIDDs.}}
\end{center}
\vspace{-5mm}
\end{figure}

In this case there are two sets of indices: The first (vertical) set
corresponds to the rows, while the second (horizontal) set corresponds
to the columns. We assign the variable name $R_i$ and $C_i$ to the row
and column index variables respectively. This distinction between the
sets of variables was originally noted in \cite{Clarke96, Bahar93}.
Figure \ref{fig:mmult} shows the QuIDD form of the above matrix
modifying a state vector via matrix-vector multiplication. Matrix
QuIDDs enjoy the same reduction rules and compression benefits as
vector QuIDDs.

\subsection{Order of Variables}
Variable ordering can drastically affect the level of compression
achieved in BDD-based structures such as QuIDDs. The CUDD package
implements sophisticated dynamic variable-reorering techniques, 
e.g., sifting, that are typically greedy in nature, but achieve significant
improvements in various applications of Binary Decision Diagrams.
However, dynamic variable reordering has significant time overhead,
and finding a good ordering in advance is preferrable in some cases.
Good variable orderings are highly dependent upon the contents of 
the decision diagram, and optimal ones are NP-hard to find.
One way to seek out an optimal ordering is to study the
problem domain. In the case of quantum computing, we notice that all
the matrices and vectors contain $2^n$ elements where $n$ is the
number of qubits represented. Additionally, the matrices are square
and non-singular \cite{NielsenC2000}. As Bahar et al. note, matrices
and vectors that do not have sizes which are a power of two require
padding with 0's \cite{Bahar93}, which can complicate real
implementations. Fortunately, no such padding is required in the realm
of quantum computing.

Hachtel et al. demonstrated that ADDs representing non-singular
matrices can be operated on efficiently with interleaved row and
column variables \cite{Hachtel96}. Interleaving implies the following
variable ordering: $R_0 \prec C_0 \prec R_1 \prec C_1 \prec ... \prec
R_n \prec C_n$. Intuitively, the interleaved ordering causes
compression to favor regularity in block sub-structures of the
matrices. Due to the fact that all matrices in the quantum domain are
non-singular, such regularity does present itself.  However, the
quantum domain offers yet another unique quality which further
enhances block pattern regularity. Not only are matrices non-singular
with power of two sizes, but they are often tensored together to allow
them to operate on multiple qubits. The tensor product $A \otimes B$
multiplies each element of $A$ by the whole matrix $B$ to create a
larger matrix which has dimensions $M_A \cdot M_B$ by $N_A \cdot
N_B$. By definition, the tensor product will propagate patterns in its
operands. As a result, our application of QuIDDs with interleaved
variable ordering scales quite nicely as the number of qubits in the
circuit increases.

It is possible to tailor the variable ordering to optimize certain
instances of circuits. For example, if a matrix is encountered that
has repeated row structures in the upper half with regular block
structure in the lower half, it could be better to define an ordering
which groups half the row variables followed by half the column
variables at the beginning (to favor row compression) and then
interleaves row and column variables halfway through (to favor block
compression).  However, as described in \ref{sec:exp}, our
experimental implementation of QuIDDs utilizes the interleaved
variable ordering for all cases due to its overall robustness in the
quantum domain.

\subsection{Matrix Multiplication}

\begin{figure}[!t]
\footnotesize
\begin{center}
\begin{tabular}{|ll|}
\hline
 & {\tt QuIDD matrix\_matrix\_multiply(QuIDD op1, QuIDD op2)} \\
 & {\tt \{ } \\
 & {\tt \hspace{1em} /* Make sure the inner dimensions (\# column } \\
 & {\tt \hspace{2em} $\hookrightarrow$ variables) agree */ } \\
{1} & {\tt \hspace{1em} if (op1.inner\_dimension != op2.inner\_dimension)} \\
 & {\tt \hspace{1em} \{ } \\
{2} & {\tt \hspace{2em} Report error; } \\
 & {\tt \hspace{1em} \} } \\
 & {\tt \hspace{1em} /* Gather all inner (column) variables */ } \\
 & {\tt \hspace{1em} /* CUDD's quasi-ring implementation requires} \\
 & {\tt \hspace{2em} $\hookrightarrow$ them. */} \\
{3} & {\tt \hspace{1em} for (int i $=$ 0; i $<$ total\_vars\_in\_system; $++$i)} \\
 & {\tt \hspace{1em} \{ } \\
{4} & {\tt \hspace{2em} if (system\_vars[i].type $==$ ``Column'')} \\
 & {\tt \hspace{2em} \{ } \\
{5} & {\tt \hspace{3em} inner\_vars.append(system\_vars[i]); } \\
 & {\tt \hspace{2em} \} } \\
 & {\tt \hspace{1em} \} } \\
 & {\tt \hspace{1em} /* Call modified version of CUDD's } \\
 & {\tt \hspace{2em} $\hookrightarrow$ Cudd\_addMatrixMultiply. This function } \\
 & {\tt \hspace{2em} $\hookrightarrow$ implements the ADD quasi-ring} \\
 & {\tt \hspace{2em} $\hookrightarrow$ multiplication but is modified to support} \\
 & {\tt \hspace{2em} $\hookrightarrow$ complex number terminals */ } \\
{6} & {\tt \hspace{1em} return\_quidd.add $=$ Cudd\_addMatrixMultiply(op1, } \\
 & {\tt \hspace{2em} op2, inner\_vars); } \\
 & {\tt \hspace{1em} /* Shift all row variables in the return ADD's} \\
 & {\tt \hspace{2em} $\hookrightarrow$ support to corresponding column } \\
 & {\tt \hspace{3em} $\hookrightarrow$ variables */ } \\
{7} & {\tt \hspace{1em} add\_support $=$ Cudd\_SupportIndex( } \\
 & {\tt \hspace{2em} $\hookrightarrow$ return\_quidd.add); } \\
{8} & {\tt \hspace{1em} for (int i $=$ 0; i $<$ total\_vars\_in\_system; $++$i)} \\
 & {\tt \hspace{1em} \{ } \\
{9} & {\tt \hspace{2em} if ((add\_support[i] $==$ 1)} \\
 & {\tt \hspace{3em} $\hookrightarrow$ \&\& (op2 is not a QuIDD vector))} \\
 & {\tt \hspace{2em} \{ } \\
{10} & {\tt \hspace{3em} Find location of corresponding column } \\
 & {\tt \hspace{4em} $\hookrightarrow$ variable and append it to } \\
 & {\tt \hspace{4em} $\hookrightarrow$ shift\_permutations; } \\
 & {\tt \hspace{2em} \} } \\
 & {\tt \hspace{1em} \} } \\
 & {\tt \hspace{1em} /* Apply shift permutations to construct a new} \\
 & {\tt \hspace{2em} $\hookrightarrow$ ADD */ } \\
{11} & {\tt \hspace{1em} return\_quidd.add $=$ Cudd\_addVectorCompose( } \\
 & {\tt \hspace{2em} $\hookrightarrow$ return\_quidd.add, shift\_permutations); } \\
{12} & {\tt \hspace{1em} return\_quidd.size $=$ op2.size; } \\
{13} & {\tt \hspace{1em} return return\_quidd; } \\
& {\tt \} } \\
\hline
\end{tabular}
\parbox{6.3in}{
\caption{ \label{fig:pseudo_mmult} Matrix multiplication for
QuIDDs. It makes use of some standard functions defined in the CUDD
package. Not shown are ``Cud\_Ref'' statements after each call to
these functions.}
}
\end{center}
\end{figure}
With the structure and variable ordering in place, operations
involving QuIDDs can now be defined. Most operations defined for ADDs
also work on QuIDDs with only slight modification. A key example is
matrix multiplication. Matrix multiplication operations with ADDs are
treated as \textit{quasi-rings} which, among other properties, means
that they have some operator $\flat$ which distributes over some
commutative operator $\sharp$ \cite{Bahar93}. This property is
critical for computing the dot-products required in matrix
multiplication, where terminal values are multiplied ($\flat$) to
produce products that are then added ($\sharp$) to create the new
terminal values of the resulting matrix.  The matrix multiplication
algorithm itself is a recursive procedure similar to the
\textit{Apply} function \cite{bryant}, but tailored to implement the
dot-product.

Another important issue in matrix multiplication is compression. To
avoid the same problem that MATLAB encounters with its ``pack''
representation, ADDs must not be decompressed to accomplish the
operation. Bahar et al. handle this by tracking the number $i$ of
``skipped'' variables between the parent (caller) and its newly
expanded child for each recursive call. A factor of $2^i$ is
multiplied by the terminal-terminal product that is reached on the
current path \cite{Bahar93}.

The primary modification that must be made when implementing this
algorithm for QuIDDs is to account for a variable ordering problem
when multiplying a matrix (operator or gate) by a vector (the qubit
state vector). A QuIDD matrix is composed of interleaved row and
column variables, whereas a QuIDD vector only depends on column
variables. If the algorithm is run as described without modification,
the resulting QuIDD vector will be composed of row instead of column
variables. The structure will be correct, but the dependence on row
variables prevents the QuIDD vector from being used in future
multiplications. Thus, we introduce a simple extension which shifts
the row variables in the new QuIDD vector to corresponding column
variables.  In other words, for each $R_i$ variable that exists in the
QuIDD vector's support, we map that variable to $C_i$. The pseudo-code
for the whole algorithm is presented in Figure \ref{fig:pseudo_mmult}.
It has worst-case time and space complexity $O(2^{2n})$, but can be
performed in $O(1)$ or $O(n)$ time and space complexity depending on
how much block regularity can be exploited in the operands. As noted
earlier, such compression is almost always achieved in the quantum
domain. The results presented in section \ref{sec:exp} verify this.

\subsection{Tensor Product and Other Operations}

\begin{figure}[!bt]
\footnotesize
\begin{center}
\begin{tabular}{|ll|}
\hline
 & {\tt QuIDD tensor(QuIDD op1, QuIDD op2)} \\
 & {\tt \hspace{1em} /* Shift all variables in op2 after op1's} \\
 & {\tt \hspace{2em} $\hookrightarrow$ variables */ } \\
{1} & {\tt \hspace{1em} add\_support = Cudd\_SupportIndex(op2) } \\
{2} & {\tt \hspace{1em} for (int i $=$ 0; i $<$ total\_vars\_in\_system; $++$i)} \\
 & {\tt \hspace{1em} \{ } \\
{3} & {\tt \hspace{2em} if (add\_support[i] $==$ 1) } \\
 & {\tt \hspace{2em} \{ } \\
{4} & {\tt \hspace{3em} Find location of next available variable} \\
 & {\tt \hspace{4em} $\hookrightarrow$ after op1's last variable and append it} \\
 & {\tt \hspace{4em} $\hookrightarrow$ to shift\_permutations;} \\
 & {\tt \hspace{2em} \} } \\
 & {\tt \hspace{1em} \} } \\
 & {\tt \hspace{1em} /* Apply shift permutations to construct a} \\
 & {\tt \hspace{2em} $\hookrightarrow$ new ADD */} \\
{5} & {\tt \hspace{1em} return\_quidd.add $=$ Cudd\_addVectorCompose(op2, } \\
 & {\tt \hspace{2em} $\hookrightarrow$ shift\_permutations); } \\
{6} & {\tt \hspace{1em} return\_quidd.size $=$ op1.size $+$ op2.size; } \\
{7} & {\tt \hspace{1em} return return\_quidd; } \\
& {\tt \} } \\
\hline
\end{tabular}
\parbox{4in}{
\caption{ \label{fig:pseudo_tensor} Tensor product for QuIDDs. 
Not shown are ``Cud\_Ref'' statements after each call to the 
CUDD functions.}
}
\end{center}
\end{figure}

The tensor product is far less complicated than matrix multiplication.
As mentioned earlier, the tensor product $A \otimes B$ produces a new
matrix which multiplies each element of $A$ by the entire matrix $B$.
Multiplication of the terminal values is easily accomplished with a
call to the recursive \textit{Apply} function with an argument that
directs \textit{Apply} to multiply when it reaches the terminals of
both operands. However, the main difficulty here lies in ensuring that
the terminals of $A$ are each multiplied by \textit{all} the terminals
of $B$. From the definition of the standard recursive \textit{Apply}
routine, we know that variables which precede other variables in the
ordering are expanded first \cite{Clarke96}. So, an algorithm must be
defined such that all of the variables in $B$ are shifted in the
current order after all of the variables in $A$ prior to the call to
\textit{Apply}. After this shift is performed, the \textit{Apply}
routine will then produce the desired behavior. \textit{Apply} starts
out with $A * B$ and expands $A$ alone until $A_{terminal} * B$ is
reached for each terminal in $A$.  Once a terminal of $A$ is reached,
$B$ is fully expanded, implying that each terminal of $A$ is
multiplied by all of $B$. Pseudo-code for this algorithm is supplied
in Figure \ref{fig:pseudo_tensor}. Notice that the size of the
resulting QuIDD is merely the sum of the two operands' sizes because
the ``size'' attribute stores the number of qubits represented by the
QuIDD, not the dimension (which is $2^{\#\ qubits}$). The time and
space complexity are based on the compression achieved in the operands
exactly as in matrix multiplication.

Other operations that must be implemented include quantum measurement
and matrix addition.  Measurement can be expressed in terms of
matrix-vector multiplication and tensor products, and so its QuIDD
implementation is merely a combination of the operations described
above. Matrix addition is easily implemented by calling \textit{Apply}
with an argument directing it to add the terminals of the
operands. Unlike the tensor product, no special variable order
shifting is required because for addition we want each terminal value
of $A$ to be added to one other terminal value of $B$, not the whole
matrix $B$.

Another interesting operation which is nearly identical to matrix
addition is terminal multiplication (so called to distinguish it from
matrix multiplication, which involves the dot-product). This algorithm
is implemented just like matrix addition except that \textit{Apply} is
directed to \textit{multiply} rather than add the terminals of the
operands. In quantum computing simulation, this operation is highly
useful when multiplying a sparse matrix like the Conditional Phase
Shift, which can be treated as a vector, by the qubit state vector.
It is computationally much cheaper to do a terminal multiplication
than represent the Conditional Phase Shift as a matrix and do
full-blown matrix multiplication.

In our implementation, QuIDD Pro, we currently support matrix
multiplication, the tensor product, matrix addition, terminal
multiplication, and scalar operations.

\section{Simulating Quantum Computation with QuIDDs}
\label{sec:exp}

This section covers the details of our experiments. We present the
quantum algorithm we simulated, a theoretical analysis of the
run-times and memory requirements, and lastly the data collected.


\subsection{Simulated Algorithm}
We tested our QuIDD data structures by simulating Grover's algorithm,
which is a quantum algorithm that offers a quadratic speed-up over
classical algorithms for searching through an unstructured database
\cite{grover}. Specifically, given $M$ items to be found in a set of
$N$ total items, the run-time of Grover's algorithm is $O \sqrt{N/M}$.

A quantum circuit representation of the algorithm involves five major
components: an \textit{oracle}, a \textit{conditional phase shift
operator}, sets of Hadamard gates, the data qubits, and an oracle
qubit. The oracle is a Boolean predicate
that acts as a filter, flipping the oracle qubit when it receives as
input an $n$ bit sequence representing the items being searched
for. In quantum circuit form, the oracle is represented as a series of
controlled NOT gates with subsets of the data qubits acting as the
control qubits and the oracle qubit receiving the action of the NOT
gates. Following the oracle, Hadamard gates put the $n$ data qubits
into a an equal superposition of all $2^n$ items in the database where
$2^n = N$. Then a sequence of gates $H^{\otimes n-1} C H^{\otimes
n-1}$, where $C$ denotes the conditional phase shift operator, are
applied iteratively to the data qubits until the probability
amplitudes of the states representing the items being searched for are
high enough to ensure successful measurement of one of them. In our
experiments, we used the tight bound formulated by Boyer et
al. \cite{Boyer96} when the number of solutions $M$ is known in
advance: $\lfloor \pi / 4 \theta \rfloor$ where $\theta =
\sqrt{M/N}$. The power of Grover's algorithm lies in the fact that the
data qubits store all $N = 2^n$ items in the database as a
superposition, allowing the oracle to ``find'' all items being
searched for \textit{simultaneously}.


\subsection{Theoretical Analysis}

Using the formulation for the bound on the number of iterations
described above, we observe that as the number of solutions $M$
decreases, the number of iterations increases \cite{Boyer96}. In terms
of the oracle, decreasing the number of solutions is achieved by
adding controls to the oracle. As an example, suppose there is some
$n$-qubit controlled NOT gate oracle $X$ that has no control qubits
initially. This oracle flips the oracle qubit for \textit{every} item
in $N$. Thus, the oracle acts as a filter accepting $n$ bit items of
any value, implying that the bit pattern it accepts is a list of don't
cares $dd...d$ with length $n$. However, when a control is added to
the NOT gate, a don't care is eliminated. For instance, if the last
data qubit for oracle $X$ becomes a $|1 \rangle$ control qubit, the
filter pattern becomes $dd...1$. Notice that the number of items being
searched for, $M$, has been cut in half and that $M$ is halved for
every control qubit that is added to the oracle. In general, the
oracles which produce the smallest $M$ induce the longest run-times
because in the formulation we chose, $M/N$ decreases as $M$ decreases.

Additionally, increasing the number of data qubits, which increases
the total number of items $N$ in the database, induces longer
run-times since $M/N$ decreases as $N$ increases. It is important to
note that increasing the number of data qubits $n$ makes $M/N$ become
\textit{exponentially} smaller because $N = 2^n$ as noted earlier.
Therefore, in our experiments the oracle which produces the longest
run-time searches for only one item ($M = 1$), and simulation will
take exponentially longer as the number of data qubits $n$ is
increased ($N = 2^n$). Since this bound on the number of iterations is
independent of the data structures used in simulation, we would expect
QuIDD-based simulations, linear algebra-based simulations, and even
real quantum computers to have run-time complexity $\Omega (2^n)$.
Empirical results for QuIDD-based vs. linear algebra-based techniques
are presented in Section \ref{sec:theory}.

As noted earlier, one of the primary advantages of quantum computers
is that they can store massive amounts of data and operate on them
in parallel. In an
equal superposition of states, $n$ qubits contain all values of $2^n$
data bits simultaneously \cite{NielsenC2000}. This implies that the
memory requirements of a real quantum computer will grow
\textit{linearly} as the number of qubits is increased in a Grover's
circuit. When simulating an instance of Grover's algorithm with the
standard linear algebra model however, the space complexity of the
simulation will grow doubly-exponentially because all values of $2^n$ data
bits are stored explicitly. QuIDDs on the other hand present a
middle-ground between real quantum computers and linear algebra
simulation in terms of space complexity. In fact, when the oracle is
polynomial in size, simulation with QuIDDs will require only polynomial
memory. To demonstrate this theoretically, we must show that when
simulating Grover's algorithm with QuIDDs the tensor product grows
polynomially for all tensored operands, and all matrix multiply
functions produce a polynomial state vector.

Bryant demonstrated that given two ROBDD structures $A$ and $B$, the
complexity of any call to \textit{Apply} with these two operands is
$O(|A| \cdot |B|)$ where $|X|$ denotes the number of nodes, including
terminals \cite{bryant}. If $A$ and $B$ are polynomial in size, the
result of the operation will also be polynomial in size. As a result,
the tensor product for QuIDDs simulating Grover's algorithm is always
polynomial as long as the qubit state vector, Hadamard, and conditional phase
shift operator are themselves all polynomial. Since each of these structures
has only two distinct elements, it can be shown by induction that the QuIDD
representations all grow polynomially.

The complexities of tensor product and matrix multiplication 
operations grow linearly during the simulating Grover's algorithm with QuIDDs.
Therefore, Grover's algorithm with a poly-size oracle requires a 
{\it polynomial} amount of memory on a classical computer. 
However, super-polynomial-size oracles may entail super-polynomial-size
vectors, even with QuIDDs because the {\it Apply}-based routines produce
super-polynomial QuIDDs when one of the arguments is super-polynomial.

\begin{table}[t]
  \begin{center}
       \begin{tabular}{|c|c|c|c|c|c|} \hline
        Ckt Size&Initial&Grover&Conditional&Oracle&Oracle\\ 
        $n$&Hadamard&Hadamard&Phase Shift&        1 &          2  \\ \hline
        20 &     80 &     83 &        21 &      99  &        108  \\ \hline
        30 &    120 &    123 &        31 &      149 &        168  \\ \hline
        40 &    160 &    163 &        41 &      199 &        228  \\ \hline
        50 &    200 &    203 &        51 &      249 &        288  \\ \hline
        60 &    240 &    243 &        61 &      299 &        348  \\ \hline
        70 &    280 &    283 &        71 &      349 &        408  \\ \hline
        80 &    320 &    323 &        81 &      399 &        468  \\ \hline
        90 &    360 &    363 &        91 &      449 &        528  \\ \hline
       100 &    400 &    403 &       101 &      499 &        588  \\ \hline
        \end{tabular} 
    \parbox{12cm}{
        \caption{\label{tab:grov_opsize} 
           Size of QuIDDs for the operators in Grover's algorithm.
                  }
    }
  \end{center}
\end{table}

\begin{table*}[t]
\scriptsize
\begin{center}
  \begin{tabular}{cc}
        \begin{tabular}{|r|c|c|c|c|c|} \hline
        \multicolumn{6}{|c|}{\bfseries Oracle 1: Runtime Comparison (s)}\\ 
        \hline\hline
        Qubits & Iters & Octave & MATLAB & Blitz++ & QuIDD Pro \\ \hline
         5     & 3     & 0.27   & 0.04   & 0       & 0.02 \\ \hline
         6     & 4     & 0.81   & 0.13   & 0       & 0.02 \\ \hline
         7     & 6     & 2.78   & 0.43   & 0.01    & 0.03 \\ \hline
         8     & 8     & 8.42   & 1.29   & 0.02    & 0.04 \\ \hline
         9     & 12    & 28.46  & 4.29   & 0.07    & 0.09 \\ \hline
        10     & 17    & 89.40  & 14.00  & 0.22    & 0.20 \\ \hline
        11     & 25    & 294.32 & 45.85  & 0.72    & 0.39 \\ \hline
        12     & 35    & 925.94 & 152.70 & 2.22    & 0.88 \\ \hline
        13     & 50    & 3086.17& 580.33 & 6.92    & 1.94 \\ \hline
        14     & 71    & 13617. & 5903.8 & 23.09   & 4.79 \\ \hline
        15     & 100   & 71000  & 59184  & 70.36   & 9.32 \\ \hline
        16     & 142   & -      & -      & 213.11  & 22.23 \\ \hline
        17     & 201   & -      & -      & 633.77  & 50.68 \\ \hline
        18     & 284   & -      & -      & 1921.11 & 112.94 \\ \hline
        19     & 402   & -      & -      & 5737.65 & 199.76 \\ \hline
        20     & 568   & -      & -      & 17420.5 & 324.51 \\ \hline
        \end{tabular}

&
        \begin{tabular}{|r|c|c|c|c|} \hline
        \multicolumn{5}{|c|}{\bfseries Oracle 1: Memory Usage Comparison (MB)}\\
         \hline\hline
        Qubits & Octave & MATLAB & Blitz++ & QuIDD Pro  \\ \hline
         5 & 0.000573 & 0.000576 & 0        & 0.00390 \\ \hline
         6 & 0.001085 & 0.000896 & 0.003906 & 0.02734 \\ \hline 
         7 & 0.002109 & 0.001536 & 0.003906 & 0.04687 \\ \hline
         8 & 0.004157 & 0.002816 & 0.007813 & 0.07031 \\ \hline
         9 & 0.008253 & 0.005376 & 0.011719 & 0.12890 \\ \hline
        10 & 0.016445 & 0.010496 & 0.019531 & 0.21094 \\ \hline
        11 & 0.032829 & 0.020736 & 0.070313 & 0.20703 \\ \hline
        12 & 0.065597 & 0.041216 & 0.074219 & 0.28125 \\ \hline
        13 & 0.131133 & 0.082176 & 0.128906 & 0.42578 \\ \hline
        14 & 0.147517 & 0.164096 & 0.25     & 0.44410 \\ \hline
        15 & 0.294973 & 0.327936 & 0.50     & 0.60547  \\ \hline
        16 & -        & -        & 1        & 0.83984  \\ \hline
        17 & -        & -        & 2        & 0.96484  \\ \hline
        18 & -        & -        & 4        & 1.58594  \\ \hline
        19 & -        & -        & 8        & 1.76562  \\ \hline
        20 & -        & -        & 16       & 2.04297  \\ \hline
        \end{tabular}
\\
\vspace{-1mm} &  \\
 {\normalsize (a)} & {\normalsize (b)} \\
 &  \\
\vspace{-1mm} &  \\
        \begin{tabular}{|r|c|c|c|c|c|} 
        \hline
        \multicolumn{6}{|c|}{\bfseries Oracle 2: Runtime Comparison (s)}\\ 
        \hline
        \hline
        Qubits & Iters & Octave   & MATLAB  & Blitz++ & QuIDD Pro \\ \hline
         12    &  25   & 544.693  & 79.31 & 1.44    &  0.45    \\ \hline
         13    &  25   & 1284.65  & 194.63  & 3.09    &  0.50    \\ \hline
         14    &  25   & 3845.06  & 827.33  & 6.88    &  0.57    \\ \hline
         15    &  25   & 12300    & 4670.8  & 15.02   &  0.57    \\ \hline
         16    &  25   & 36156    & 37805  & 32.33   &  0.68    \\ \hline
        \end{tabular}
&
        \begin{tabular}{|r|c|c|c|c|} \hline
        \multicolumn{5}{|c|}{\bfseries Oracle 2: Memory Usage Comparison (MB)}\\
        \hline\hline
        Qubits &  Octave   & MATLAB   & Blitz++ & QuIDD Pro \\ \hline
        12     &  0.065597 & 0.041216 & 0.20312 & 0.23047  \\ \hline
        13     &  0.131133 & 0.082176 & 0.32422 & 0.31250  \\ \hline
        14     &  0.147517 & 0.164096 & 0.57422 & 0.33594  \\ \hline
        15     &  0.294973 & 0.327936 & 1.07422 & 0.37109  \\ \hline
        16     &  0.589885 & 0.655616 & 2.07422 & 0.40234  \\ \hline
        \end{tabular}
 \\
\vspace{-1mm}
  &  \\
 {\normalsize (c)} & {\normalsize (d)} \\
  \end{tabular}
  \parbox{6in} {
         \vspace{-2mm}
         \caption{\label{tab:resources} Resource utilization of our 
          simulations of Grover's algorithm with two oracles.
                 }
                }
  \end{center}
         \vspace{-5mm}
\end{table*}

\subsection{Experimental Results}
We implemented QuIDD Pro as a C++ program which utilizes the CUDD
library \cite{cudd} to handle the underlying ADD structure with row
and column variables interleaved.  So far, we have only used QuIDD Pro
in a few types of simulations to verify that QuIDDs result in a useful
amount of compression for practical applications.  One such simulation
involves running Grover's algorithm \cite{grover} for identification
of keys in an unstructured database of all $2^n$ possible values of an
$n$-qubit register. We compare this to an optimized implementation of
Grover's algorithm using Blitz++, a high-performance numerical linear algebra
library for C++ \cite{blitz}, MATLAB, and Octave, a mathematical package 
similar to MATLAB.

As shown in Table \ref{tab:grov_opsize}, the sizes of QuIDDs
representing operators used in Grover's algorithm do grow linearly
with the size of the system as we expected. For a given number of
qubits in Grover's circuit, a change in the problem requires only a
change in the oracle and it is this that governs time and space
complexity per-iteration.  To demonstrate the effectiveness of QuIDDs
we include results from simulations on various oracles for different
system-sizes.

The first oracle we use identifies the key consisting of all $1$'s.
This case therefore has a single solution in the search space of size
$2^{n-1}$. All computations were made on an AMD 1.2GHz dual processor
machine with 1GB RAM. Table \ref{tab:resources}a contrasts the
runtimes of different computational packages simulating this instance
of Grover's algorithm for various numbers of qubits. QuIDD Pro runs
much faster than the other packages. We believe this is due to the
fact that the size of the QuIDD data structures are so small compared
to the full-blown matrices and vectors in the other packages and
therefore require fewer computations. Notice that MATLAB and Octave
exhibit prohibitive runtimes as the number of qubits in the circuit
increases, most likely due to the fact that they are interpreted
rather than compiled languages like Blitz++ and QuIDD Pro. As a result,
we only included data for MATLAB and Octave for up to $15$ qubits.

Table \ref{tab:resources}b contrasts the memory utilization for
all the packages. Notice that MATLAB, Octave, and Blitz++ start to
exhibit noticeable exponential growth from $14$ to $15$ qubits. In
Blitz++, the exponential growth continues up to $20$ qubits, while
QuIDD Pro grows only linearly as we expected.

The second oracle we use identifies keys that are a specified value
{\it modulo} 1024. We ran this case for all values between 0 and 1023
for various number of qubits. All computations were made on an Intel
Xeon 2GHz dual processor machine with 1GB RAM. Tables
\ref{tab:resources}c and \ref{tab:resources}d respectively
show the runtime and memory requirements for various tools simulating
this case. Again, QuIDD Pro clearly demonstrates the best performance
in terms of runtime and memory. This oracle effectively checks the 10
least significant qubits for equality with the specified value. Since
the number of solutions, $M$ and the size of the search space, $N$
grow by a factor of 2 with each additional qubit, the number of
Grover-iterations remains invariant. Runtime is now governed purely by
the size of the system making this oracle useful in demonstrating the
asymptotics. We ran QuIDD Pro with this oracle to $25$ qubits and using
linear least-squares regression we find the memory (MB) to vary as
 $7.5922 + 0.0410n$.


To verify that the Boyer et al. formulation \cite{Boyer96} resulted in
the exact number of Grover iterations to generate the highest
probability of measuring the items being searched for, we tracked
their probabilities as a function of the number of iterations. For
this experiment, we used four different oracles, each with $11$,$12$,
and $13$ qubit circuits. The first oracle is called ``Oracle N'' and
represents an oracle in which all the data qubits act as controls to
flip the oracle qubit (this oracle is equivalent to Oracle 1 in the
last subsection). The other oracles are ``Oracle N-1'', ``Oracle
N-2'', and ``Oracle N-3'', which all have the same structure as Oracle
N minus $1$,$2$, and $3$ controls, respectively. As described earlier,
each removal of a control doubles the number of items being searched
for in the database. For example, Oracle N-2 searches for $4$ items in
the data set because it recognizes the bit pattern $111...dd$.

\begin{table}[t]
\small
  \begin{center}
    \begin{tabular}{|c|c|c|c|} \hline
      Oracle & 11 Qubits & 12 Qubits & 13 Qubits  \\ \hline
      $N$ & 25 & 35 & 50 \\ \hline
      $N-1$ & 17 & 25 & 35 \\ \hline
      $N-2$ & 12 & 17 & 25 \\ \hline
      $N-3$ & 8 & 12 & 17 \\ \hline
    \end{tabular}
    
    \parbox{4.6in}{
      \caption{
        \label{tab:g_iters}
    Number of Grover iterations at which Boyer et al. \cite{Boyer96} predict 
    the highest probability of measuring one of the items sought.
      }
    }
  \end{center}
\end{table}

\begin{figure}[!ht]
  \begin{center}
    \begin{tabular}{cc}
         \includegraphics[width=7.5cm]{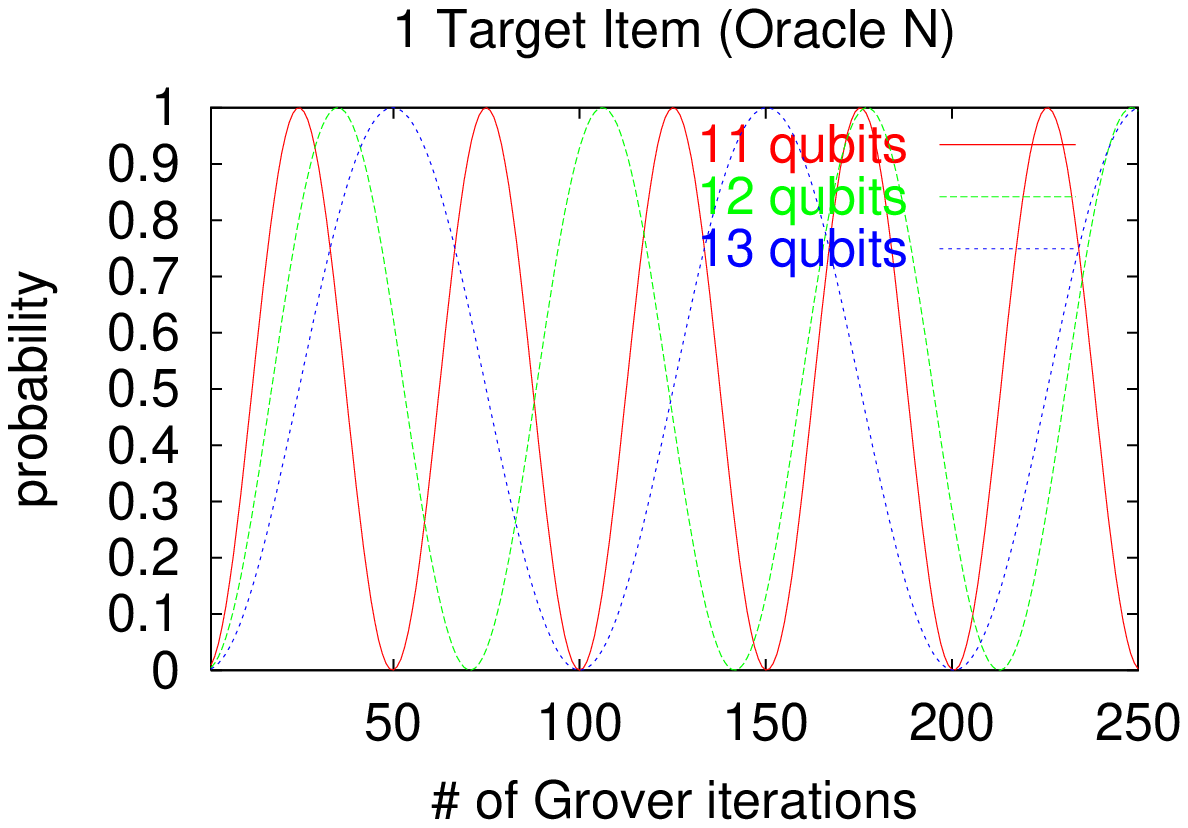} &
         \includegraphics[width=7.5cm]{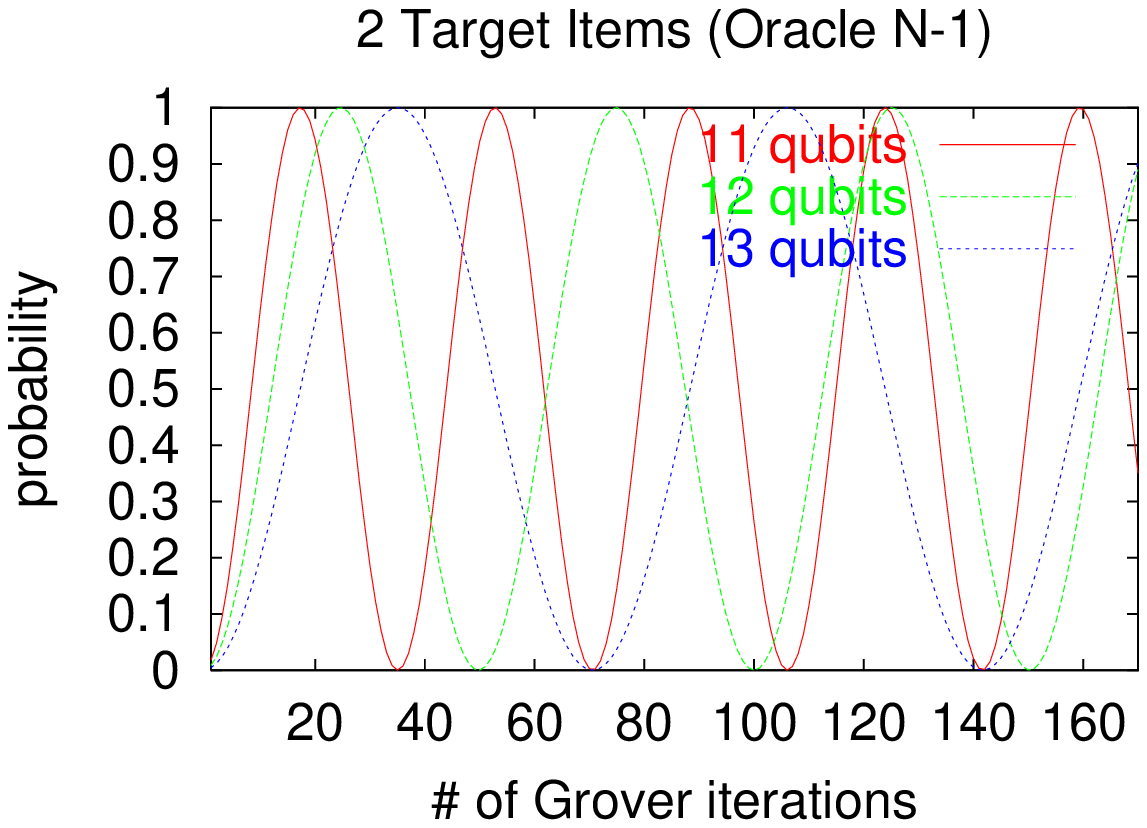} \\
         \includegraphics[width=7.5cm]{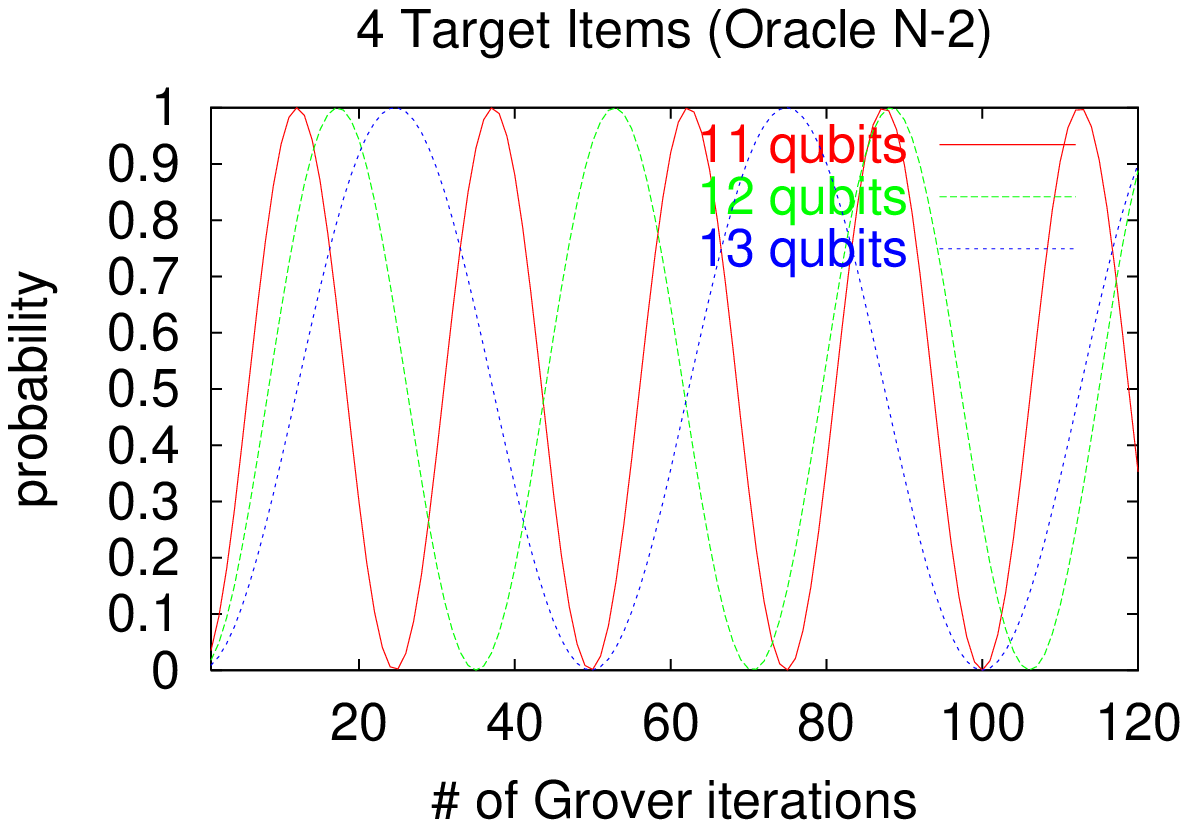} &
         \includegraphics[width=7.5cm]{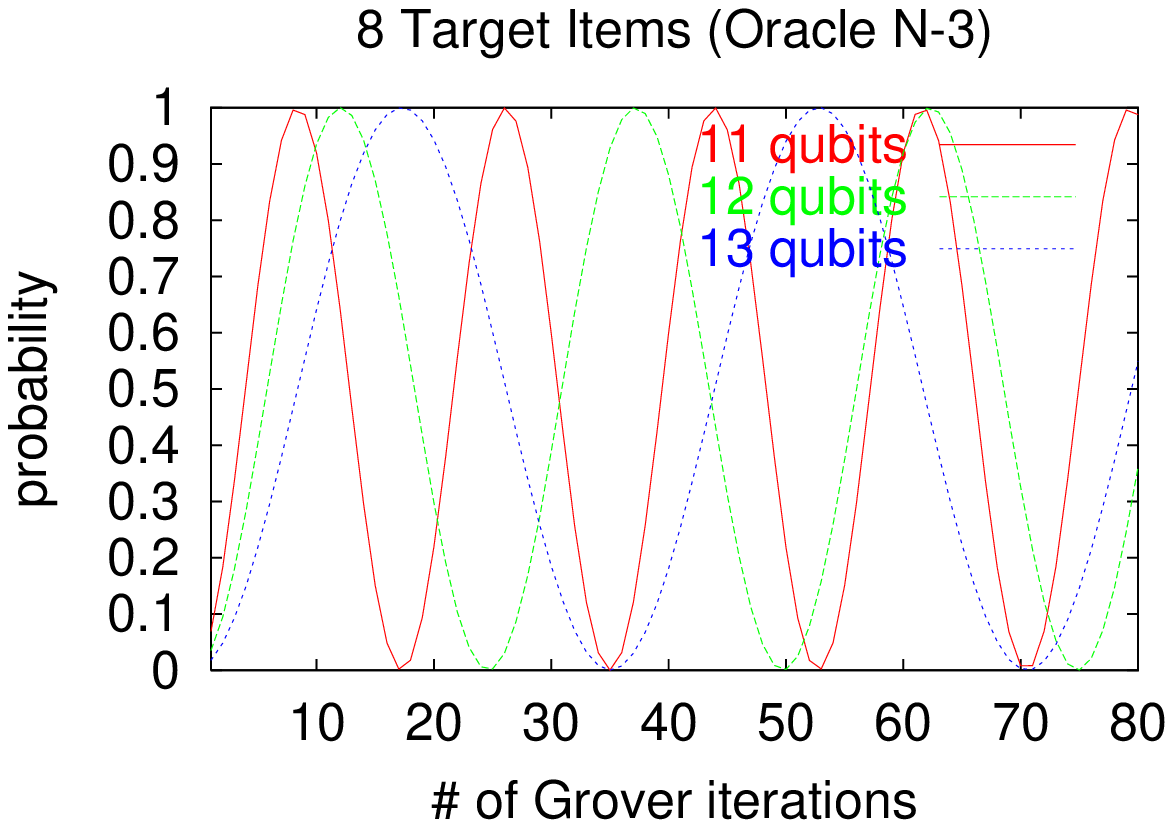}
    \end{tabular}

    \parbox{6in}{
      \caption{
        \label{fig:prob}
        Probability of successful search for one, two, four
        and eight items as a function of the number of iterations
        after which the measurement is performed (11, 12 and 13 qubits).
      }
    }
  \end{center}
\end{figure}

Table \ref{tab:g_iters} shows the optimal number of iterations
produced with the Boyer et al. formulation for all the instances
tested. Figure \ref{fig:prob} plots the probability successfully finding
any of the items sought against the number of Grover iterations.
In the case of Oracle N, we plot the probability of measuring the
single item being searched for. Similarly, for Oracles N-1, N-2, and
N-3, we plot the probability of measuring any one of the $2$, $4$, and
$8$ items being searched for, respectively. By comparing results in 
Table \ref{tab:g_iters} with those in Figure \ref{fig:prob}, 
it can be easily verified that Boyer et al. correctly predict
the number of iterations at which measurement is most likely
to produce items sought.

\section{Conclusions and Future Work}
\label{sec:conclusions}

We believe that QuIDDs provide a practical and highly efficient
approach to high-performance quantum computational simulation. Their
key advantage over straightforward matrix-based simulations 
is much faster execution and lower memory usage. 
We developped new software package, QuIDD Pro, for high-performance
simulation of quantum circuits.  To this end, we simulated Grover's
search algorithm by performing respective gate operations in QuIDD Pro.
Our results indicate that QuIDD Pro achieves exponential
memory savings compared to MATLAB and other competitors.
Additionally, QuiDD Pro was asymptotically faster than competing simulators,
however not quite as fast as Grover's algorithm. Empirically, QuIDD Pro
spends on the order of $1.66^q$ time searching a system with $q$ qubits.
Grover's search executed on a quantum computer requires time on the order
of $\sqrt{2^q}\approx 1.41^q$, not counting various potential overhead.
Since MATLAB-based simulations explicitly store large matrices,
their space complexity must be $\Omega(4^q)$, and the same is true
for time complexity because every matrix element is computed and used.
Simulations using Blitz++ required $\Theta(2^q)$ memory which was
dominated by the dense state-vector, and their runtime must grow at least
as fast. Yet, the simulation using QuIDD Pro required only a linearly-growing
amount of memory.

We note that in our experiments, numerical precision of complex-valued
terminals was fixed. QuiDDPro currently incorporates variable-precision
integer and floating-point number types, and all reported experiments
were performed with sufficient numerical precisious to avoid round-off
errors. However, we have not yet experimented with precision requirements 
at larger number of qubits and the effects of round-off errors on 
our simulations. That will be addressed in our future work
that also includes simulating other quantum algorithms,
such as Shor's \cite{Shor97}, and modelling the effects of errors
and decoherence in quantum computation. 



\end{document}